\documentclass[amssymb,amsmath,aps,pre,showpacs,preprint]{revtex4}
\usepackage{amssymb}

\usepackage{graphicx}

\begin{document}

\title{Properties of non-FCC hard-sphere solids predicted by density functional theory. }
\author{James F. Lutsko}
\affiliation{Center for Nonlinear Phenomena and Complex Systems, Universit\'{e} Libre de
Bruxelles, C.P. 231, Blvd. du Triomphe, 1050 Brussels, Belgium}
\pacs{64.10.+h, 05.20.-y, 05.70.Ce}
\date{\today }

\begin{abstract}
The free energies of the FCC, BCC, HCP and Simple Cubic phases for hard spheres are calculated
as a function of density using the Fundamental Measure Theory models of Rosenfeld et al (PRE 55, 4245 (1997)),
 Tarazona (PRL 84, 694 (2001)) and Roth et al (J. Phys.: Cond. Matt. 14,  12063 (2002)) in the Gaussian approximation.
 For the FCC phase,
the present
work confirms the vanishing of the Lindemann parameter (i.e. vanishing of the width of the Gaussians) near close packing
 for all three models and 
the results for the HCP phase are nearly identical. For the BCC phase and for  packing fractions above  $\eta \sim 0.56$, all three theories show multiple solid structures differing in the widths of the Gaussians. In all three cases, one of 
these structures shows the expected vanishing of the Lindemann parameter at close packing, but this physical structure
is only thermodynamically favored over the unphysical structures in the Tarazona theory and even then, some unphysical behavior persists at lower densities. The simple cubic phase is stabilized in the model of Rosenfeld et al. for a range of densities and in the Tarazona model only very near close-packing. 
\end{abstract}
\maketitle

\section{Introduction}

The Fundamental Measure Theory (FMT) approach to building approximate free
energy density functionals has proven very successful in describing the
properties of inhomogeneous hard-sphere fluids including the hard-sphere
FCC\ solid\cite{Rosenfeld1,Rosenfeld2,Rosenfeld_1997_1,tarazona_2000_1,tarazona_2002_1,WhiteBear}. The original form
of FMT\ proposed by Rosenfeld\cite{Rosenfeld1,Rosenfeld2} was based
on a generalization of the ideas underlying scaled particle theory. The
Rosenfeld functional gave a useful description of an inhomogeneous fluid but
did not stabilize the FCC solid. One important formal property of the
functional was that its second functional derivative with respect to the
density reduced to the Percus-Yevik direct correlation function in the
uniform liquid limit. Subsequent refinements were motivated in particular by
the requirement that the free energy functional reduce to the known exact
form when the density was restricted to zero- and one-dimensional systems%
\cite{Rosenfeld_1997_1}. The resulting theories, discussed in more detail
below, retains all of the useful properties of the original Rosenfeld
functional while also stabilizing the FCC solid and predicting the
liquid-solid coexistence reasonably accurately\cite{Rosenfeld_1997_1,tarazona_2000_1,tarazona_2002_1,WhiteBear}. The theory also
gives a good description of the mean-squared displacement of the atoms in
the solid phase including the nontrivial prediction that the mean-squared
displacement goes to zero as the density approaches close packing.

The purpose of this paper is to present the results of the application of
the FMT to other crystalline structures of the hard-sphere solid. There are
two good reasons for carrying out such a study. First, from a practical
point of view, one would like to be able to use the FMT to study solid-solid
phase transitions as well as a basis for the study, via thermodynamic
perturbation theory, of non-FCC solids along the lines discussed in
refs.\cite{CurtinAschroftPert,Lutsko_1991_3,Lutsko_1992_1, Song_Perturbation_Theory}. To do so with any confidence first requires that the predictions
of the FMT for non-FCC hard-sphere solids be well understood. The second
reason is that more loosely-packed crystal structures may provide a more
demanding test of the theory than does the FCC phase since the structure of
the FCC phase is not that dissimilar to that of the liquid:\ other structures
differ more from the liquid structure, e.g. in terms of the nearest-neighbor
coordination. As shown below, this expectation is borne out and the
predictions of the properties of the BCC\ phase in particular are not as
satisfactory as for the FCC phase.

The remainder of this paper is organized as follows. Section \ref{sec2} reviews
the elements of Density Functional Theory (DFT) and FMT and discusses the difference between the three
FMTs considered in this work. The results of the calculations are presented
in Section \ref{sec3} where it is shown that the theories give very different 
results for different lattice structures. The final Section summarizes
the results of these calculations and discusses obstacles that prevent
any simple modification of these theories to give a better description of the 
BCC phase.

\section{Fundamental Measure Theory}
\label{sec2}

As is usual in DFT, the Helmholtz free energy $F$ is written as a
sum of an ideal contribution $F_{id}$ and an excess contribution, $F_{ex}$.
The former is given by%
\begin{equation} \label{fid}
\beta F_{id}\left[ \rho \right] =\int \left[ \rho \left( 
\overrightarrow{r}\right) \ln \Lambda ^{3}\rho \left( \overrightarrow{r}%
\right) -\rho \left( \overrightarrow{r}\right) \right] d\overrightarrow{r}
\end{equation}%
where $\rho \left( \overrightarrow{r}\right) $ is the local density, $N=\int
\rho \left( \overrightarrow{r}\right) d\overrightarrow{r}$ is the total
number of particles and $\beta =1/k_{B}T$ is the inverse temperature. 
In
the FMT\ approximation, the excess free energy functional is written in
terms of a set of local functions%
\begin{equation}
\beta F_{ex}\left[ \rho \right] = \int \sum_{i=1}^{3}\beta \phi
_{i}\left( \left\{ n_{\alpha }\left( \overrightarrow{r},\left[ \rho \right]
\right) \right\} \right) d\overrightarrow{r}  \label{fex}
\end{equation}%
which, as indicated, include a functional dependence on the density through
a family of local functionals of the form%
\begin{equation}
n_{\alpha}\left( \overrightarrow{r},\left[ \rho \right]\right) =\int w_{\alpha }\left( 
\overrightarrow{r}-\overrightarrow{r}^{\prime }\right) \rho \left( 
\overrightarrow{r}^{\prime }\right) d\overrightarrow{r}^{\prime }.  \label{2}
\end{equation}%
The weights $w_{\alpha}\left(\overrightarrow{r}\right)$ occurring in eq.(\ref{2}) are $\Theta \left( \frac{\sigma }{2}%
-r\right) $, $\delta \left( r-\frac{\sigma }{2}%
\right) $ , $\widehat{r}\delta \left( r-\frac{%
\sigma }{2}\right) $ and $\widehat{r}\widehat{r}%
\delta \left( r-\frac{\sigma }{2}\right) $ yielding the local density functionals
 which will be denoted $\eta \left( \overrightarrow{r}\right) $, $s\left( 
\overrightarrow{r}\right) $, $\overrightarrow{v}\left( \overrightarrow{r}%
\right) $, $\overleftrightarrow{T}\left( \overrightarrow{r}\right) $
respectively. For a uniform liquid, in which $\rho \left( \overrightarrow{r}%
\right) =\overline{\rho }$, one has, in three dimensions, that $\eta \left( 
\overrightarrow{r}\right) =\frac{1}{6}\pi \sigma ^{3}\overline{\rho }$ which
is the usual definition of the packing fraction. The names of the other
quantities are motivated by their scalar, vector and tensor natures
respectively. The first two contributions to the excess free energy are%
\begin{eqnarray}
\beta \phi _{1} &=&-\frac{1}{\pi \sigma ^{2}}s\left( \overrightarrow{r}\right) \ln
\left( 1-\eta \left( \overrightarrow{r}\right) \right)   \label{3} \\
\beta \phi _{2} &=&\frac{1}{2\pi \sigma }\frac{s\left( \overrightarrow{r}\right)
^{2}-v^{2}\left( \overrightarrow{r}\right) }{\left( 1-\eta \left( 
\overrightarrow{r}\right) \right) }  \nonumber
\end{eqnarray}%
These are the same as in the original Rosenfeld theory. The third function
has been the focus of most efforts to refine the FMT and we write it in the
form%
\begin{equation}
\beta \phi _{3}=\frac{1}{8\pi \left( 1-\eta \left( \overrightarrow{r}\right)
\right) ^{2}}f\left( \overrightarrow{r}\right) 
\end{equation}%
where the only fundamental constraint on \ $f$ is that $\lim_{\rho
\rightarrow 0}f=1$ \cite{Rosenfeld_1997_1}. Three proposals in the literature, aside from the
original form $f=\frac{1}{3}s^{3}\left( \overrightarrow{r}\right) $, which
does not stabilize the bulk FCC solid, are%
\begin{eqnarray} \label{f3}
f_{RSLT} &=&\frac{1}{3}s^{3}\left( \overrightarrow{r}\right) \left( 1-\xi
^{2}\left( \overrightarrow{r}\right) \right) ^{3},\;\xi ^{2}=\frac{%
v^{2}\left( \overrightarrow{r}\right) }{s^{2}\left( \overrightarrow{r}%
\right) } \\
f_{T} &=&\frac{3}{2}\left( \overrightarrow{v}\left( \overrightarrow{r}%
\right) \cdot \overleftrightarrow{T}\left( \overrightarrow{r}\right) \cdot 
\overrightarrow{v}\left( \overrightarrow{r}\right) -s\left( \overrightarrow{r%
}\right) v^{2}\left( \overrightarrow{r}\right) -Tr\left( \overleftrightarrow{%
T}^{3}\left( \overrightarrow{r}\right) \right) +s\left( \overrightarrow{r}%
\right) Tr\left( \overleftrightarrow{T}^{2}\left( \overrightarrow{r}\right)
\right) \right)   \nonumber \\
f_{E} &=&\frac{2}{3}\left( 1-\eta \left( \overrightarrow{r}\right) \right)
^{2}\left( \beta f_{ex}^{l}\left( \eta \left( \overrightarrow{r}\right)
\right) +\log \left( 1-\eta \left( \overrightarrow{r}\right) \right) -\frac{%
3\eta \left( \overrightarrow{r}\right) }{1-\eta \left( \overrightarrow{r}%
\right) }\right) f_{T}  \nonumber
\end{eqnarray}%
where $f_{RSLT}$ was proposed by Rosenfeld et al\cite{Rosenfeld_1997_1} $%
f_{T}$ is the tensor form proposed by Tarazona\cite{tarazona_2000_1} and $%
f_{E}$ is a heuristic modification of $f_{T}$ that allows for the insertion
of an empirical equation of state for the liquid, $\beta f_{ex}^{l}\left(
\eta \right) $. In the latter case, if the equation of state is chosen to be
the Carnahan-Starling equation\cite{HansenMcdonald}, the resulting free
energy functional is the single-species version of the so-called ''White
Bear'' functional of Roth et al.\cite{WhiteBear}, which was also proposed
by\ Tarazona\cite{tarazona_2002_1}. The first and second forms are 
closely related. In fact,\ writing the tensor density as $%
\overleftrightarrow{T}\left( \overrightarrow{r}\right) =\overleftrightarrow{U%
}\left( \overrightarrow{r}\right) +\frac{1}{3}s\left( \overrightarrow{r}%
\right) \overleftrightarrow{1}$, where $\overleftrightarrow{U}\left( 
\overrightarrow{r}\right) $ is traceless, a natural approximation for the
traceless part is to make it proportional to the only traceless tensor that
can be formed from the other densities%
\begin{equation}
\overleftrightarrow{U}\left( \overrightarrow{r}\right) \approx A\frac{%
\overrightarrow{v}\left( \overrightarrow{r}\right) \overrightarrow{v}\left( 
\overrightarrow{r}\right) -\frac{1}{3}\overrightarrow{v}^{2}\left( 
\overrightarrow{r}\right) \overleftrightarrow{1}}{s\left( \overrightarrow{r}%
\right) }
\end{equation}%
where $A$ is a constant and the denominator is needed so that $%
\overleftrightarrow{U}\left( \overrightarrow{r}\right) $ scales linearly
with the density in the sense that $\rho \left( \overrightarrow{r}\right)
\rightarrow \lambda \rho \left( \overrightarrow{r}\right) $ implies that $%
\overleftrightarrow{U}\left( \overrightarrow{r}\right) \rightarrow \lambda 
\overleftrightarrow{U}\left( \overrightarrow{r}\right) $. Substituting this
into $f_{T}$ gives $f_{T}=\frac{1}{3}s^{3}\left( \overrightarrow{r}\right)
\left( 1-3\xi ^{2}+3A\xi ^{4}-A^{3}\xi ^{6}\right) $ which reduces to $%
f_{RSLT}$ provided that one takes $A=1$. The choice $A=0$ corresponds to the
original FMT\ theory of Rosenfeld. Note that one could also divide by $\left| \overrightarrow{v}\left( \overrightarrow{r}\right)\right|$
rather than $s\left( \overrightarrow{r}\right)$ in this ansatz, which  is equivalent
to the replacement $A \rightarrow A' \xi^{-1}\left( \overrightarrow{r}\right)$
giving $f_{T}=\frac{1}{3}s^{3}\left( \overrightarrow{r}\right)
\left( 1-3\xi ^{2}+A'(3-A'^{2})\xi ^{3} \right)$. Taking $A'=1$ gives the ``interpolation''
form for $\phi_{3}$ discussed in ref. \cite{Rosenfeld_1997_1}.

The calculations reported here were performed using  the standard approximation of the density as a
sum of Gaussians%
\begin{equation}
\rho \left( \overrightarrow{r}\right) =\sum_{n}f\left( \overrightarrow{r}-%
\overrightarrow{R}_{n}\right) 
\end{equation}%
where $f\left( r\right) =x_{0}\left( \frac{\alpha }{\pi }\right) ^{3/2}\exp
\left( -\alpha r^{2}\right) $. For a perfect lattice, one has occupancy $x_{0}=1$
whereas values $x_{0}<1$ characterize a solid with vacancies. The parameter $%
\alpha $ characterizes the width of the Gaussians and $\left\{ 
\overrightarrow{R}_{n}\right\} $ is the set of lattice vectors of the
desired crystal structure. The density can also be written in terms of components in
reciprocal space as 
\begin{equation}
\rho \left( \overrightarrow{r}\right) =\sum_{n}\exp \left( i\overrightarrow{K%
}_{n}\cdot \overrightarrow{r}\right) \rho _{n}
\end{equation}%
where $\left\{ \overrightarrow{K}_{n}\right\} $ is the corresponding set of
reciprocal lattice vectors and $\rho _{n}=x_{0}N_{cell}\rho _{cell}\exp
\left( -K_{n}^{2}/4\alpha \right) $ with $N_{cell}$ being the number of
lattice sites per unit cell and $\rho _{cell}$ being the number of unit cells per
unit volume. Thus, $N_{cell}\rho _{cell}$ is just the number density of the
perfect solid. From the fundamental theorems of DFT, the equilibrium values
of the parameters $x_{0},\rho _{cell}$ and $\alpha $ are those which
minimize the grand potential, $\beta \Omega = \beta F_{id}\left[ \rho \right] +\beta F_{ex}%
\left[ \rho \right] -\beta \mu N$
 for fixed chemical potential $\mu 
$ and volume $V$. The  number of atoms is $N = \overline{\rho}V$ where the average density is $\overline{\rho} = \frac{1}{V}\int_{V} \rho(\overrightarrow{r})d\overrightarrow{r} = x_{0}N_{cell}\rho_{cell}$.

The spatial integrals in equations (\ref{fid})-(\ref{fex}) were performed by
 sampling the arguments on a grid of points covering the standard unit cell. 
The primary consideration in fixing the number of points in the grid is that 
the spacing be sufficiently fine so as to be able to sample the width of the 
Gaussians, $\left(\Delta R\right)^{2} = \frac{3}{2\alpha}$. In the calculations discussed below, 
the
 spacing of the points in each Cartesian direction was set to of $dr = a/(2q)$
 where $a$ is the lattice parameter and $q$ was initially set to 20 and then doubled until $dr/\Delta R < 0.5$. Increasing the number of points beyond this limit had no significant effect on the results. Note that in
the Gaussian approximation and in the limit of $\alpha a^2 >> 1$ , the ideal contribution to the free energy approaches
the asymptotic value
\begin{equation} 
\beta f_{id} \rightarrow x_{0}\left(\frac{3}{2}\ln\left(\frac{x_{0} \alpha \sigma^{2}}{\pi}\right) -2.5 \right).
\end{equation}
This approximation was used for $\alpha a^2 > 200$. The plots of the free energy
as a function of $\alpha$ presented below were all generated based on sampling the
free energy at intervals of $\delta \alpha \sigma^2 = 1, 5, 10, 50$ and $ 250$ for 
$\alpha \sigma^{2} < 20, 200, 2000, 10,000$ and $20,000$, 
respectively. The free energy minima were determined using the the Simplex algorithm of Nelder 
and Mead, see e.g. ref. \cite{NR}, as
implemented in the Gnu Scientific Library\cite{GSL}, which was terminated when the simplex size was smaller
than $10^{-4}$.

\section{Results}
\label{sec3}

\subsection{The FCC crystal}

\begin{figure}[tbp]
\begin{center}
\resizebox{12cm}{!}{
{\includegraphics{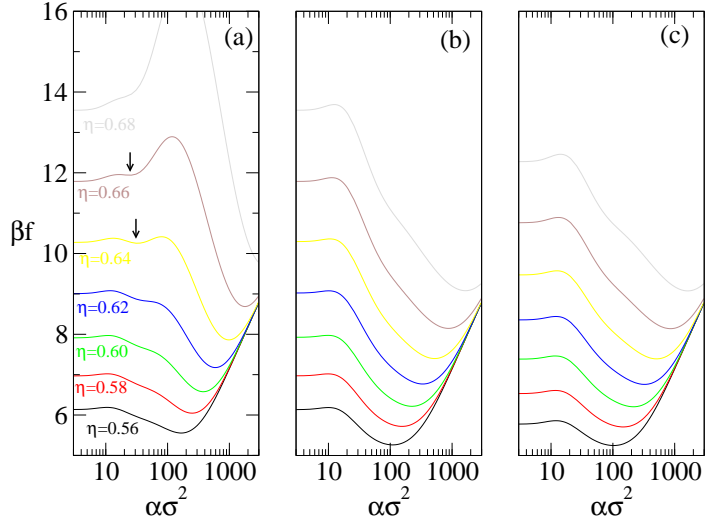}}}
\end{center}
\caption{Dimensionless Helmholtz free energy per atom, $\beta f = \beta F/N $,  as a function of $\alpha$ for the FCC hard-sphere solid as calculated
using (a) the RSLT theory, (b) the Tarazona theory and (c) the White Bear theory. From bottom to 
top, the curves are for $\overline{\eta} = 0.56 - 0.68$ in steps of $0.02$. The arrows indicate
the positions of the secondary solid minima.}
\label{fig1}
\end{figure}

Figure \ref{fig1} shows the free energies as a function of $\alpha $ for the FCC\
solid as calculated from the three theories for a variety of average packing
fractions $\overline{\eta} =\frac{\pi }{6}\overline{\rho }\sigma ^{3}$. These
calculations are performed for $x_{0}=1$ which is known to be correct at the
solid minimum\cite{Rosenfeld_1997_1} and which we have verified is very
close to the free energy minimum at all values of $\alpha $. All three
theories show a minimum for all densities at $\alpha =0$ corresponding to
the liquid phase. This is unphysical for the higher densities which are well
above random close packing, $\overline{\eta} _{rcp}\sim 0.64$, and is a result of the
use of the Percus-Yevik and Carnahan-Starling approximate equations of state
which are only singular at $\overline{\eta} =1$. On the other hand, note the
consistency in the location of the free energy minimum at finite $\alpha $
as well as of the value of the free energy at the minimum. While the
location of the minimum in RSLT theory is somewhat higher than for the other
two theories, the differences are not large.  For all three theories, the
location of the solid minimum increases rapidly with density which is a
characteristic feature of the FMT\cite{Rosenfeld_1997_1,tarazona_2002_1}. The Tarazona and White Bear functionals differ primarily
in the small-$\alpha $ region, due to the difference in the equations of
state of the liquid, but at large $\alpha $ the differences are negligible.
The main difference between the tensor theories  and the RSLT\ theory is also in the small-$\alpha$ region where, for $\overline{\eta} > 0.64$ the RSLT\ theory predicts much larger
energy barriers between the liquid and solid minima than do the tensor
theories. Note also that for a narrow range of densities, $0.61\precsim 
\overline{\eta }\precsim 0.63$, the RSLT\ theory actually shows \emph{two}
minima at non-zero values of $\alpha $. This appears to be quite unphysical
since all atoms are, by hypothesis, assumed to be identical and so it is
hard to see how there could be two metastable states with the same density but different vibrational amplitudes. 

The equilibrium free energy of the solid phase as a function of average
density was determined by minimizing the free energy functional with respect
to both $\alpha $ and the lattice parameter. From this, the densities of the
coexisting liquid and solid as predicted by each theory were determined in
the usual way by finding liquid and solid states with the same pressure and
chemical potential. The RSLT theory gives the coexisting liquid and solid at
packing fractions $0.491$ and $0.534$ respectively (compared to $%
0.491$ and $0.540$ reported in \cite{Rosenfeld_1997_1} and $0.491$ and $0.534$ reported in \cite{Song_FMT}), the Tarazona theory
gives $0.466$ and $0.516$ (compared to $0.467$ and $0.516$ reported in ref.%
\cite{tarazona_2002_1}) and the White Bear theory $0.489$ and $0.535$
(compared to $0.489$ and $0.536$ reported in ref.\cite{tarazona_2002_1}).
For comparison, the values from simulation are $0.492$ and $0.545$\cite%
{Hoover_1968_1}. The reason that the Tarazona theory gives relatively poor
predictions for coexistence is, paradoxically, because it gives a rather
good description of the solid while the liquid equation of state is still
that of the  Percus-Yevik theory which is inaccurate at the high densities
that characterize coexistence\cite{tarazona_2002_1}. This is one of the main motivations for
introducing the empirical equation of state so that a free energy functional
is obtained that yields accurate values over the entire range of structures.

\begin{figure}[tbp]
\begin{center}
\resizebox{12cm}{!}{
{\includegraphics{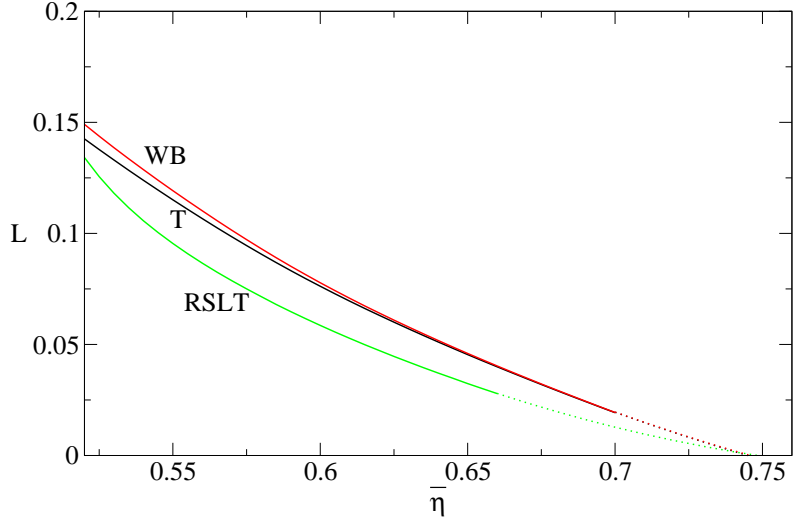}}}
\end{center}
\caption{The Lindemann parameter for the FCC phase as a function of $\overline{\eta}$ as calculated using  the RSLT theory (lower line) , the Tarazona theory (middle curve labeled T) and the White Bear theory (upper curve labeled WB). Also shown as dotted lines are the quadratic interpolation of the curves to $L=0$  based on the data for $\overline{\eta} > 0.60$. }
\label{fig2}
\end{figure}

The predicted
Lindemann parameter, $L$ which is the ratio of the root-mean-square displacement 
to the nearest neighbor distance, is
shown in Fig.\ref{fig2}. For the FCC crystal, it decreases monotonically with increasing 
density and extrapolates to zero at a density near the FCC close-packing 
density of $\overline{\eta}_{FCC-CP} = \pi\sqrt{2}/6 \sim 0.74$.

It is worth noting that similar calulations have been performed for the closely related HCP phases. In all cases, the
results closely mirrored those for the FCC phase shown here. In particular, the free 
energies were always larger than, but within 1\% of, the FCC free energy.

\subsection{The BCC crystal}

\begin{figure}[tbp]
\begin{center}
\resizebox{12cm}{!}{
{\includegraphics{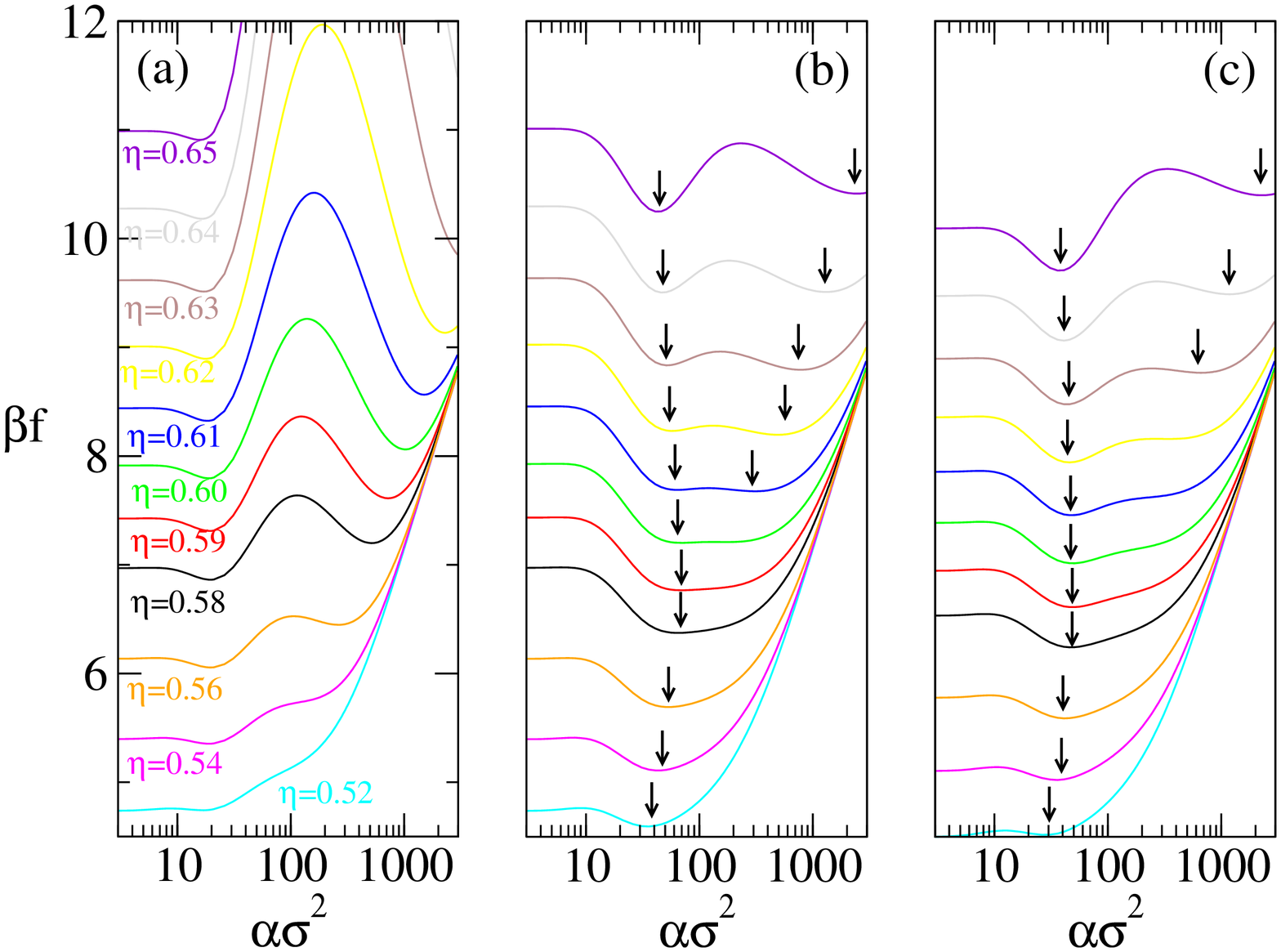}}}
\end{center}
\caption{Same as \ref{fig1} for the BCC phase. In (b) and (c), the arrows show the positions of the solid minima. From bottom to 
top, the curves are for $\overline{\eta} = 0.52, 0.54, 0.56, 0.58, 0.59, 0.60, 0.61, 0.62, 0.63, 0.64, 0.65$.}
\label{fig3}
\end{figure}

Figure \ref{fig3} shows the same calculations for the BCC solid. In this case, all
three theories give multiple solid minima over some range of densities. For $%
\overline{\eta} \geq 0.56,$the RSLT theory gives two solid minima with the low-density
minimum having lower free energy than the high-density minimum and thus
being the preferred state. The Tarazona theory shows multiple minima for $%
\overline{\eta }\geq 0.61$ with energies that differ by less than one
percent. In the range $0.61\leq \overline{\eta }\leq 0.63$ the low-$\alpha $
branch has slightly lower free energy while for $0.635\leq \overline{\eta }$
the high-$\alpha $ minimum is favored. The White Bear functional only
develops a high-$\alpha $ minimum for $0.63\leq \overline{\eta }$ and in all
cases the low-$\alpha $ minimum has lower free energy.

\begin{figure}[tbp]
\begin{center}
\resizebox{12cm}{!}{
{\includegraphics{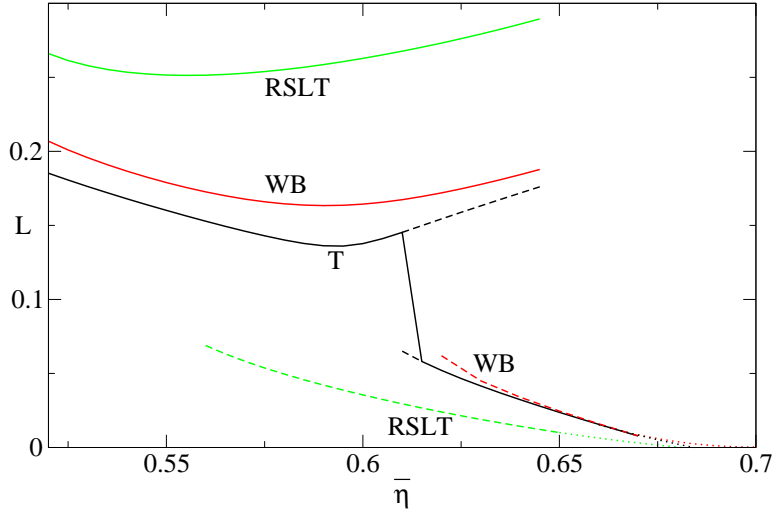}}}
\end{center}
\caption{The Lindemann parameter for the BCC phase as a function of packing fraction, $\overline{\eta}$, as calculated using  the RSLT theory, the Tarazona theory (labeled T) and the White Bear theory (labeled WB). Both the low-$\alpha$ and high-$\alpha$ branches are shown with the stable branch being drawn with full lines and the unstable branch with dashed lines. Also shown as dotted lines are the quadratic interpolation of the curves to $L=0$  based on the data for $\overline{\eta} > 0.60$. }
\label{fig4}
\end{figure}

As seen in Fig. \ref{fig4}, the behavior of the Lindemann parameter for the BCC crystal is more complex than in the case of the 
FCC phase. All three theories give a ``low-$\alpha$'' branch 
which is not monotonic: the Lindemann parameter decreases with increasing density 
at low densities, but then increases with increasing density at high densities. 
This behavior is similar to that of the effective-liquid-type DFTs\cite{Lutsko_1990_1, tarazona_2000_1}
and is considered quite unphysical since, if the trend  persisted 
up to close packing, it would 
imply that spheres are able to interpenetrate. Rather, one expects the mean-squared displacement to 
decrease uniformly with increasing density as in the case of the FCC crystal.
All three FMTs also give  a second,
 ``high-$\alpha$'' branch which does behave physically at high densities and, in particular, the Lindemann parameter 
extrapolates to zero near the BCC close packing density of $\overline{\eta}_{BCC-CP} = \pi\sqrt{3}/8 \sim 0.68$. However, for with the RSLT theory,  up 
to $\overline{\eta} = 0.625$, the well-behaved branch has higher free energy than does the unphysical branch. The Tarazona theory gives similar results except that the high-$\alpha$ minima are thermodynamically favored at high density so that, as noted by Tarazona\cite{tarazona_2000_1, tarazona_2002_1}, the prediction of the theory is that the Lindemann parameter appears to vanish near close-packing, as it should. 
Nevertheless, the prediction of the theory is that the Lindemann parameter increases with increasing density at intermediate densities as well as giving a discontinuous jump in the Lindemann parameter for a packing fraction near $0.60$, all of which seem somewhat unphysical. Finally, the White Bear theory also has a second branch but it only appears at very high densities and also has higher free energy than does the low-$\alpha$ branch. Comparison of Figures 2b and 2c shows that the reason for the observed behavior is that the Carnahan-Starling equation of state used in the White Bear functional lowers the liquid free energy relative to  the Percus-Yevik equation of state that comes out of the Tarazona theory. As $\alpha$ increases, the effect lessens until at high $\alpha$ the theories are almost identical. Thus, the free energy of the low-$\alpha$ minimum is somewhat lower in the White Bear theory leading to its stability persisting even up to high densities. 

\subsection{The SC phase}

\begin{figure}[tbp]
\begin{center}
\resizebox{12cm}{!}{
{\includegraphics{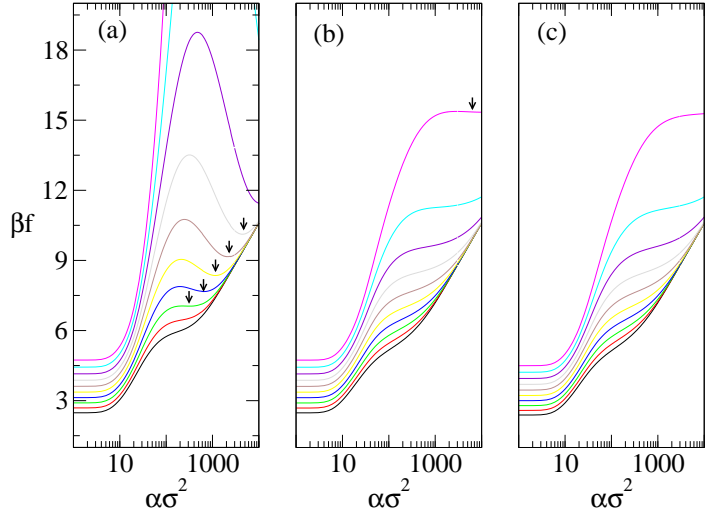}}}
\end{center}
\caption{Same as Fig. \ref{fig1} for the SC phase. From bottom to 
top, the curves are for $\overline{\eta}$ in the range $[0.43,0.52]$ in steps of $0.01$.}
\label{fig5}
\end{figure}

For a purely repulsive potential, the simple cubic phase is expected to be unstable with respect
to shear. Nevertheless, there is no reason that DFT should not give a well-defined free energy
for this structure since it should be metastable. Indeed, at least one of the effective-liquid
theories, the GELA, does give a stable structure with small values of $\alpha$\cite{Lutsko_1990_1}.

Figure \ref{fig5} shows the free energy curves calculated using the three theories. The results
from the RSLT theory are much as expected: the simple cubic phase stabilizes at quite low densities, $\overline{\eta} \sim 0.45$, and
the location of the free energy minimum increases with increasing density and appears to diverge as
the simple cubic close packing, $\overline{\eta}_{SC-CP} = \pi/6 \sim 0.524$ is approached. In this case, there is no indication
of multiple minima. In contrast, the tensor theories show no sign of a $\alpha \ne 0$ minimum except at densities very close to close packing. Extending the calculations up to $\alpha \sigma^{2} = 20,000$, no minimum was found in the $\overline{\eta} = 0.51$ curve using either of the tensor models while for $\overline{\eta} = 0.52$ the Tarazona theory stablizes the solid at $\alpha \sigma^{2} \sim 11,500$ and there is again no minimum using the White Bear theory.

\subsection{Relative Stability}
\label{rel}

\begin{table}[tbp]
\caption{The minimum packing fractions at which the various FMT models predict the 
different crystal structures become stable and the approximate value of the Gaussian
parameter $\alpha$ at this density. }
\label{tab1}%
\begin{ruledtabular}
\begin{tabular}{cccc}
Structure & Theory & $\overline{\eta }_{sol}$ & $\alpha \sigma^{2}$ \\ \hline
FCC & RSLT & 0.49 & 27 \\
 & T & 0.46 & 22 \\
 & WB & 0.48 & 26 \\
BCC & RSLT & 0.51 & 14 \\
 & T & 0.48 & 20 \\
 & WB & 0.51  & 24 \\
SC & RSLT &  0.47 & 1150 \\
SC & T &  0.52 & 11,250\\
SC & WB &  --- & --- \\
\end{tabular}
\end{ruledtabular}
\end{table}

The last question we adress is the relative stability of the various structures. Table \ref{tab1} shows the minimum densities at which each structure becomes stable in the various FMT models. For the FCC and BCC structures, the RSLT and WB theories give quite similar results while the Tarazona theory generally stabilizes the solid at a lower density about 5\% below the other theories. In all cases, the FCC phase is stable at lower densities than the BCC phase. 

\begin{figure}[tbp]
\begin{center}
\resizebox{12cm}{!}{
{\includegraphics{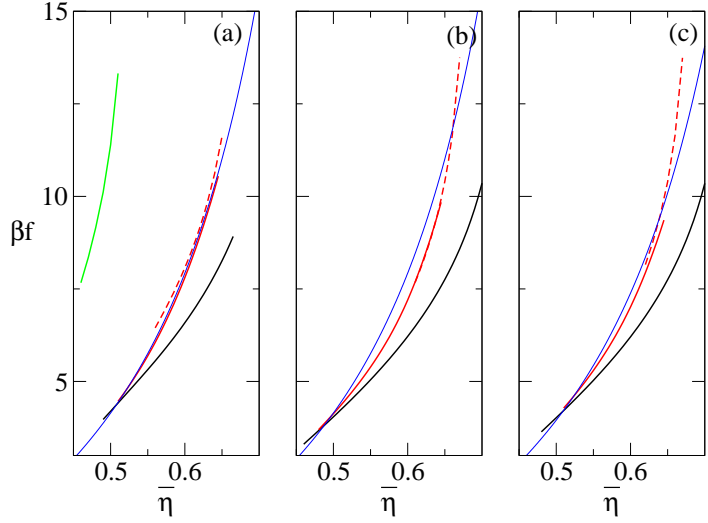}}}
\end{center}
\caption{The free energy minima for the liquid phase and the various crystal structures
as functions of the average packing fraction calculated
using (a) the RSLT theory, (b) the Tarazona theory and (c) the White Bear theory. The lowest
curves are for the FCC structure, the intermediate curves are the low-$\alpha$ (full) and 
high-$\alpha$ (dashed) BCC branches and the light curve that spans the width of the figures
are for the liquid phase. In (a), the upper curve is for the simple cubic structure.}
\label{fig6}
\end{figure}

The free energies as functions of the average density are shown in Fig. \ref{fig6}. In all models, the FCC phase has the lowest free energy of any solid phase at all densities. The free energy of the BCC phase is very close to that of the FCC at low densities and steadily diverges at increasing densities. The FCC phase has slightly higher free energy than the liquid at the lowest densities. The low-$\alpha$ branch of the BCC phase always has lower free energy than the liquid, although in the RSLT theory they are almost identical. In the RSLT, the simple cubic structure has much higher free energy than any other phase.

\section{Conclusions}
\label{sec4}

In summary, the present calculations confirm the fact that the three FMT's discussed all give reasonable results for the FCC phase at all densities up to FCC close-packing with the only exception being the appearance of a spurious low-$\alpha$ minimum at high densities in the RSLT theory. On the other hand, when applied to the BCC phase, these results indicate several unphysical properties of the Fundamental Measure Theory functionals which must be carefully considered when using them, e.g., as the basis for thermodynamic perturbation theory. All of the theories appear to give reasonable results at moderate densities, say up to $\overline{\eta} \sim 0.56$.  However, the simplest model, the RSLT, gives multiple solid minima for all packing fractions in the BCC phase above about $0.56$. The Tensor theories do not give multiple minima for the FCC phase but they do for the BCC phase. As the density increases, the stable minima in the Tarazona theory switches from the unphysical branch to the physically well-behaved branch whereas the White Bear theory always picks out the unphysical branch. Only the RSLT and Tarazona  theories stablize the simple cubic phase, at least for $\alpha \sigma^{2} < 20,000$, but  the Lindemann parameter is, in this case, well-behaved and there are no spurious minima. Unfortunately for the goal of choosing a single theory to describe both the liquid and all solid phases, the RSLT and WB theories give poor results for the BCC phase Lindemann parameter while the Tarazona theory gives the worst quantitative description of FCC-liquid coexistence.

It is interesting to speculate on the resolution of the common deficiencies in these theories. It seems reasonable to attribute the unphysical behavior at high densities, which is manifested by the fact that the unphysical low-$\alpha$ minimum is favored, to a poor description of the liquid phase: the free energy of the liquid must clearly diverge at high densities and in fact it has long been debated whether the hard-sphere liquid-phase equation of state possesses a singularity at high densities\cite{McCoy_2006_1} - at, e.g., random close packing\cite{Ma_1985} or FCC close packing\cite{Sanchez_1_1994}. 
If the free energy of the liquid did indeed diverge at random close packing, then the curves shown in Figs. \ref{fig1} and \ref{fig3} would be deformed with the low-$\alpha$ free energies ``pulled up''. This could conceivable eliminate altogether the low-$\alpha$ minima and thus give a reasonable picture of the BCC phase. However, as appealing as this scenario is, it cannot be implemented naively using eq.(\ref{f3}) to give the equivalent of the ``White Bear'' improvement to the Tarazona theory. This is because eq.(\ref{f3}) requires that the equation of state be evaluated at the local packing fraction $\overline{\eta}(\overrightarrow{r})$ and for the solid phase there are always points at which $\eta(\overrightarrow{r})$ is very nearly equal to one unless $\alpha$ is very small. (In particular,  it is easy to see that $\lim_{\alpha \rightarrow \infty}\eta(\overrightarrow{0}) = 1$.) Thus, simply inserting an equation of state with a divergence for $\overline{\eta} < 1$ is not an option since this will lead to divergent free energies in the solid phase, even for the already well-described FCC phase.

One approximation not tested here is the use of the Gaussian profile for the densities. Tarazona has performed calculations using more general parameterizations of the density and he reports that the generalization beyond the Gaussian is of negligble imortance for the equation of state\cite{tarazona_2000_1}. Groh and Mulder have also supplied evidence for the FCC phase that the Gaussian approximation is quite accurate, especially near the peaks of the density distribution\cite{Groh_Mulder}. This suggests that the effects reported here are not due to the Gaussian approximation but definitive proof will require further calculations using non-Gaussian profiles.  

In summary, the Fundamental Measure Theories work well for the FCC phase at all densities while for the BCC phase, the FMTs are best applied to the solid phase for densities near liquid-solid coexistence. While they can in principle be used to model the BCC phase at higher densities, and while they do show some desirable features at high density like the vanishing of the Lindemann parameter at close packing, they also suffer from the same problem that plagued earlier DFT models: namely, the unphysical behavior of the Lindemann parameter for the BCC phase at intermediate densities. On the other hand, the SC phase is well modeled, at least using the RSLT theory, perhaps because it only exists at relatively low densities.   

\begin{acknowledgments}
I am grateful to Marc Baus and Xeyeu Song for several stimulating discussions on this topic.
This work was supported in part by the European Space Agency/PRODEX
under contract number C90238.
\end{acknowledgments}

\bigskip 

\bibliographystyle{apsrev}

\bibliography{../physics}

\begin{thebibliography}{20}
\expandafter\ifx\csname natexlab\endcsname\relax\def\natexlab#1{#1}\fi
\expandafter\ifx\csname bibnamefont\endcsname\relax
  \def\bibnamefont#1{#1}\fi
\expandafter\ifx\csname bibfnamefont\endcsname\relax
  \def\bibfnamefont#1{#1}\fi
\expandafter\ifx\csname citenamefont\endcsname\relax
  \def\citenamefont#1{#1}\fi
\expandafter\ifx\csname url\endcsname\relax
  \def\url#1{\texttt{#1}}\fi
\expandafter\ifx\csname urlprefix\endcsname\relax\def\urlprefix{URL }\fi
\providecommand{\bibinfo}[2]{#2}
\providecommand{\eprint}[2][]{\url{#2}}

\bibitem[{\citenamefont{Rosenfeld}(1989)}]{Rosenfeld1}
\bibinfo{author}{\bibfnamefont{Y.}~\bibnamefont{Rosenfeld}},
  \bibinfo{journal}{Phys. Rev. Lett.} \textbf{\bibinfo{volume}{63}},
  \bibinfo{pages}{980} (\bibinfo{year}{1989}).

\bibitem[{\citenamefont{Rosenfeld et~al.}(1990)\citenamefont{Rosenfeld,
  Levesque, and Weis}}]{Rosenfeld2}
\bibinfo{author}{\bibfnamefont{Y.}~\bibnamefont{Rosenfeld}},
  \bibinfo{author}{\bibfnamefont{D.}~\bibnamefont{Levesque}}, \bibnamefont{and}
  \bibinfo{author}{\bibfnamefont{J.-J.} \bibnamefont{Weis}},
  \bibinfo{journal}{J. Chem. Phys.} \textbf{\bibinfo{volume}{92}},
  \bibinfo{pages}{6818} (\bibinfo{year}{1990}).

\bibitem[{\citenamefont{Rosenfeld et~al.}(1997)\citenamefont{Rosenfeld,
  Schmidt, L{\"o}wen, , and Tarazona}}]{Rosenfeld_1997_1}
\bibinfo{author}{\bibfnamefont{Y.}~\bibnamefont{Rosenfeld}},
  \bibinfo{author}{\bibfnamefont{M.}~\bibnamefont{Schmidt}},
  \bibinfo{author}{\bibfnamefont{H.}~\bibnamefont{L{\"o}wen}}, ,
  \bibnamefont{and} \bibinfo{author}{\bibfnamefont{P.}~\bibnamefont{Tarazona}},
  \bibinfo{journal}{Phys. Rev. E} \textbf{\bibinfo{volume}{55}},
  \bibinfo{pages}{4245} (\bibinfo{year}{1997}).

\bibitem[{\citenamefont{Tarazona}(2000)}]{tarazona_2000_1}
\bibinfo{author}{\bibfnamefont{P.}~\bibnamefont{Tarazona}},
  \bibinfo{journal}{Phys. Rev. Lett.} \textbf{\bibinfo{volume}{84}},
  \bibinfo{pages}{694} (\bibinfo{year}{2000}).

\bibitem[{\citenamefont{Tarazona}(2002)}]{tarazona_2002_1}
\bibinfo{author}{\bibfnamefont{P.}~\bibnamefont{Tarazona}},
  \bibinfo{journal}{Physica A} \textbf{\bibinfo{volume}{306}},
  \bibinfo{pages}{243} (\bibinfo{year}{2002}).

\bibitem[{\citenamefont{Roth et~al.}(2002)\citenamefont{Roth, Evans, Lang, and
  Kahl}}]{WhiteBear}
\bibinfo{author}{\bibfnamefont{R.}~\bibnamefont{Roth}},
  \bibinfo{author}{\bibfnamefont{R.}~\bibnamefont{Evans}},
  \bibinfo{author}{\bibfnamefont{A.}~\bibnamefont{Lang}}, \bibnamefont{and}
  \bibinfo{author}{\bibfnamefont{G.}~\bibnamefont{Kahl}}, \bibinfo{journal}{J.
  Phys.: Cond. Matt.} \textbf{\bibinfo{volume}{14}}, \bibinfo{pages}{12063}
  (\bibinfo{year}{2002}).

\bibitem[{\citenamefont{Curtin and Ashcroft}(1986)}]{CurtinAschroftPert}
\bibinfo{author}{\bibfnamefont{W.~A.} \bibnamefont{Curtin}} \bibnamefont{and}
  \bibinfo{author}{\bibfnamefont{N.~W.} \bibnamefont{Ashcroft}},
  \bibinfo{journal}{Phys. Rev. Lett.} \textbf{\bibinfo{volume}{56}},
  \bibinfo{pages}{2775} (\bibinfo{year}{1986}).

\bibitem[{\citenamefont{Lutsko and Baus}(1991)}]{Lutsko_1991_3}
\bibinfo{author}{\bibfnamefont{J.~F.} \bibnamefont{Lutsko}} \bibnamefont{and}
  \bibinfo{author}{\bibfnamefont{M.}~\bibnamefont{Baus}}, \bibinfo{journal}{J.
  Phys.: Condens. Matter} \textbf{\bibinfo{volume}{3}}, \bibinfo{pages}{6547}
  (\bibinfo{year}{1991}).

\bibitem[{\citenamefont{Tejero et~al.}(1992)\citenamefont{Tejero, Lutsko,
  Colot, and Baus}}]{Lutsko_1992_1}
\bibinfo{author}{\bibfnamefont{C.~F.} \bibnamefont{Tejero}},
  \bibinfo{author}{\bibfnamefont{J.~F.} \bibnamefont{Lutsko}},
  \bibinfo{author}{\bibfnamefont{J.~L.} \bibnamefont{Colot}}, \bibnamefont{and}
  \bibinfo{author}{\bibfnamefont{M.}~\bibnamefont{Baus}},
  \bibinfo{journal}{Phys. Rev. A} \textbf{\bibinfo{volume}{46}},
  \bibinfo{pages}{3373} (\bibinfo{year}{1992}).

\bibitem[{\citenamefont{Warshavsky and Song}(2004)}]{Song_Perturbation_Theory}
\bibinfo{author}{\bibfnamefont{V.~B.} \bibnamefont{Warshavsky}}
  \bibnamefont{and} \bibinfo{author}{\bibfnamefont{X.}~\bibnamefont{Song}},
  \bibinfo{journal}{Phys. Rev. E} \textbf{\bibinfo{volume}{69}},
  \bibinfo{pages}{061113} (\bibinfo{year}{2004}).

\bibitem[{\citenamefont{Hansen and McDonald}(1986)}]{HansenMcdonald}
\bibinfo{author}{\bibfnamefont{J.-P.} \bibnamefont{Hansen}} \bibnamefont{and}
  \bibinfo{author}{\bibfnamefont{I.}~\bibnamefont{McDonald}},
  \emph{\bibinfo{title}{Theory of Simple Liquids}}
  (\bibinfo{publisher}{Academic Press}, \bibinfo{address}{San Diego, Ca},
  \bibinfo{year}{1986}).

\bibitem[{\citenamefont{Press et~al.}(1993)\citenamefont{Press, Teukolsky,
  Vetterling, and Flannery}}]{NR}
\bibinfo{author}{\bibfnamefont{W.~H.} \bibnamefont{Press}},
  \bibinfo{author}{\bibfnamefont{S.~A.} \bibnamefont{Teukolsky}},
  \bibinfo{author}{\bibfnamefont{W.~T.} \bibnamefont{Vetterling}},
  \bibnamefont{and} \bibinfo{author}{\bibfnamefont{B.~P.}
  \bibnamefont{Flannery}}, \emph{\bibinfo{title}{Numerical Recipes in C}}
  (\bibinfo{publisher}{Claredon Press}, \bibinfo{address}{Oxford},
  \bibinfo{year}{1993}).

\bibitem[{GSL()}]{GSL}
\emph{\bibinfo{title}{The gnu scientific library}},
  \eprint{http://sources.redhat.com/gsl}.

\bibitem[{\citenamefont{Warshavsky and Song}(2006)}]{Song_FMT}
\bibinfo{author}{\bibfnamefont{V.~B.} \bibnamefont{Warshavsky}}
  \bibnamefont{and} \bibinfo{author}{\bibfnamefont{X.}~\bibnamefont{Song}},
  \bibinfo{journal}{Phys. Rev. E} \textbf{\bibinfo{volume}{73}},
  \bibinfo{pages}{031110} (\bibinfo{year}{2006}).

\bibitem[{\citenamefont{Hoover and Ree}(1968)}]{Hoover_1968_1}
\bibinfo{author}{\bibfnamefont{W.~G.} \bibnamefont{Hoover}} \bibnamefont{and}
  \bibinfo{author}{\bibfnamefont{F.~H.} \bibnamefont{Ree}},
  \bibinfo{journal}{J. Chem. Phys.} \textbf{\bibinfo{volume}{49}},
  \bibinfo{pages}{3609} (\bibinfo{year}{1968}).

\bibitem[{\citenamefont{Lutsko and Baus}(1990)}]{Lutsko_1990_1}
\bibinfo{author}{\bibfnamefont{J.~F.} \bibnamefont{Lutsko}} \bibnamefont{and}
  \bibinfo{author}{\bibfnamefont{M.}~\bibnamefont{Baus}},
  \bibinfo{journal}{Phys. Rev. E} \textbf{\bibinfo{volume}{41}},
  \bibinfo{pages}{6647} (\bibinfo{year}{1990}).

\bibitem[{\citenamefont{Clisby and McCoy}(2006)}]{McCoy_2006_1}
\bibinfo{author}{\bibfnamefont{N.}~\bibnamefont{Clisby}} \bibnamefont{and}
  \bibinfo{author}{\bibfnamefont{B.~M.} \bibnamefont{McCoy}},
  \bibinfo{journal}{J. Stat. Phys.} \textbf{\bibinfo{volume}{122}},
  \bibinfo{pages}{15} (\bibinfo{year}{2006}).

\bibitem[{\citenamefont{Ma and Ahmadi}(1986)}]{Ma_1985}
\bibinfo{author}{\bibfnamefont{D.}~\bibnamefont{Ma}} \bibnamefont{and}
  \bibinfo{author}{\bibfnamefont{G.}~\bibnamefont{Ahmadi}},
  \bibinfo{journal}{J. Chem. Phys.} \textbf{\bibinfo{volume}{84}},
  \bibinfo{pages}{3449} (\bibinfo{year}{1986}).

\bibitem[{\citenamefont{Sanchez}(1994)}]{Sanchez_1_1994}
\bibinfo{author}{\bibfnamefont{I.~C.} \bibnamefont{Sanchez}},
  \bibinfo{journal}{J. Chem. Phys.} \textbf{\bibinfo{volume}{101}},
  \bibinfo{pages}{7003} (\bibinfo{year}{1994}).

\bibitem[{\citenamefont{Groh and Mulder}(2000)}]{Groh_Mulder}
\bibinfo{author}{\bibfnamefont{B.}~\bibnamefont{Groh}} \bibnamefont{and}
  \bibinfo{author}{\bibfnamefont{B.}~\bibnamefont{Mulder}},
  \bibinfo{journal}{Phys. Rev. E} \textbf{\bibinfo{volume}{61}},
  \bibinfo{pages}{3811} (\bibinfo{year}{2000}).

\end{thebibliography}

\end{document}